\documentclass[10pt, twoside]{article}

\usepackage{graphics}
\usepackage{amsmath,amssymb,amsfonts}
\usepackage{comment}
\usepackage{multirow,bigdelim}
\usepackage{varwidth}
\usepackage{fancyvrb}
\usepackage[colorlinks,citecolor=blue,urlcolor=blue,linkcolor=blue,bookmarks=false]{hyperref}
\usepackage[mode=buildnew]{standalone}
\usepackage{psfrag,xspace}
\usepackage{etoolbox}
\usepackage[usenames,dvipsnames]{xcolor}
\usepackage[sort]{natbib}
\usepackage{authblk}

\usepackage{tikz}

\def\code#1{\texttt{#1}}

\def\ind{\perp\!\!\!\perp}

\newcommand{\var}{\textit{var}}

\newcommand{\Qb}{\mathbb{Q}}
\newcommand{\Pb}{\mathbb{P}}

\newcommand{\Qn}{\mathbb{Q}_N}

\newcommand{\E}{\mathbb{E}}

\newcommand{\bY}{\mathbf{Y}}
\newcommand{\by}{\mathbf{y}}
\newcommand{\bX}{\mathbf{X}}
\newcommand{\bZ}{\mathbf{Z}}
\newcommand{\bx}{\mathbf{x}}
\newcommand{\bz}{\mathbf{z}}

\newcommand{\bzero}{\mathbf{0}}

\DeclareSymbolFont{bbold}{U}{bbold}{m}{n}
\DeclareSymbolFontAlphabet{\mathbbold}{bbold}
\newcommand{\one}{\mathbbold{1}}

\newtheorem{assumption}{Assumption}

\newtheorem{remark}{Remark}

\begin{document}

\title{drpop: Efficient and Doubly Robust Population Size Estimation in R}
  \author[1]{Manjari Das}
  \author[1]{Edward H. Kennedy}
  \affil[1]{Department of Statistics \& Data Science,
  Carnegie Mellon University, Pittsburgh PA}
  \date{}
  \maketitle
  \thispagestyle{empty}
  
\begin{abstract}
  This paper introduces the {R} package {drpop} to flexibly estimate total population size from incomplete lists. Total population estimation, also called capture-recapture, is an important problem in many biological and social sciences. A typical dataset consists of incomplete lists of individuals from the population of interest along with some covariate information. The goal is to estimate the number of unobserved individuals and equivalently, the total population size. {drpop} flexibly models  heterogeneity using the covariate information, under the assumption that two lists are conditionally independent given covariates. This can be a much weaker assumption than full marginal independence often required by classical methods. Moreover, it can incorporate complex and high dimensional covariates, and does not require parametric models like other popular methods. In particular,  our estimator is doubly robust and has fast convergence rates even under flexible non-parametric set-ups. {drpop} provides the user with the flexibility to choose the model for estimation of intermediate parameters and returns the estimated population size, confidence interval and some other related quantities. In this paper, we illustrate the applications of {drpop} in different scenarios and we also present some performance summaries.
\end{abstract}


\section{Introduction}

One crucial step in working with capture-recapture or population size estimation problem, is applying the appropriate identification assumption. Population size estimation is inherently a missing data problem, and hence, one requires some kind of assumption to ensure that the population size is identifiable from the observed data. One should maintain caution while making identifying assumptions, since it can induce bias if not valid for the data \citep{you2021estimation, tilling2001capture, hook1999recommendations, link2003nonidentifiability, huggins2001note}. To ensure identifiability, in general, all approaches use some lack of dependence assumption among the lists. The simplest approach works with two lists assuming marginal independence \citep{petersen1896yearly}. Some advances in this stream include \citet{schnabel1938estimation, Darroch1958indeplists, burnham1979robust} and \citet{lee1994estimating}. In the presence of covariates, one can use mild assumptions to ensure identifiability. \citet{tilling1999capture, huggins1989statistical, das2021doubly} among others assumed that two lists are independent conditional on the covariate and presented non-parametric estimators. This conditional independence assumption is milder than the marginal independence assumption. This assumption can be used for a wide range of data collection scenarios.

And following that, the next step is to account for any heterogeneity present in the data. Real data is often far from homogeneous. Unmodelled or wrongly modelled heterogeneity can also lead to misleading inference \citep{link2003nonidentifiability, carothers1973effects}. To account for heterogeneity and/or list dependence, some of the literature used intricate data structures, e.g., complex covariate information. These approaches are mostly model-based. To name a few, there are \citet{link2003nonidentifiability, carothers1973effects, Fienberg1972loglinear, tilling1999capture, pollock2002use, huggins1989statistical, alho1993estimating, yip2001unified}. Capture probabilities of individuals, i.e., probability of being observed, are often non-linear or complex functions of the covariates \citep{huggins2007non, stoklosa2012robust} and estimation using linear or strong parametric models might lead to bias. \citet{stoklosa2012robust} has presented a generalized additive model approach to address this issue. 

{drpop} implements the doubly robust estimators of capture probability and population size from \citet{das2021doubly}, which rely on assuming two lists are only conditionally rather than marginally independent. These methods are flexible yet efficient, with small mean squared error even in non-parametric models involving continuous or high-dimensional covariates. 

\subsection{Existing packages and softwares}
There are several {R} packages and other softwares available for capture-recapture data. Table \ref{tab:packagelist} shows a list of some of the existing {R} packages along with the new {drpop}. Some of the existing packages are designed for improving estimation and runtime for the classical set-up whereas, others are primarily designed for open population and/or continuous time captures. In the open population set-up, the population is not fixed. There can be addition or deletion. When the population is fixed over the duration of data collection, then it is called a closed population set-up. For this paper, we will focus only on the closed population set-up with discrete capture times. Discrete capture times is the same as a finite number of lists. This set-up generally holds for data collected over a shorter time period.

One of the oldest softwares is {MARK} \citep{white1999program, white2001advanced} (extended to {R} with package {RMark} by \citet{laake2008rmark}) and it works on both closed and open population set-ups. For the closed population, it uses the conditional likelihood approach of \citet{huggins1989statistical, huggins1991some} incorporating individual covariate information. {Rcapture} \citep{JSSv019i05} uses log-linear approach for closed population set-ups implementing the work of \citet{cormack1989log, rivest2004loglinear, rivest2007applications, rivest2001improved, cormack1985examples, cormack1991inference, frischer1993estimating}. It does not use covariate information but models heterogeneity using lists information. \citet{chao2014capture, chao2001applications} presented the {R} package {CARE1} that is designed mainly for closed human populations and uses sample coverage approach. It does not use covariate information either. One of the most recent packages is {VGAM} \citep{yee2015vgam}. It is designed for closed population and uses conditional likelihood method while also using covariate information to model heterogeneity. One of the main advantages of {VGAM} is the ability to model the heterogeneity as non-linear functions of the covariates using vector generalized linear and additive models.

There are other existing softwares and packages, for example, software {M-Surge} \citet{choquet2004m}, and packages like {mra} \citep{mcdonald2018package}, {marked} \citep{laake2013marked}, {multimark} \citep{mcclintock2015multimark}. These mainly focus on a broader variety of capture-recapture problems, like open population and continuous time captures which are beyond the scope of this paper. For a detailed review and performance comparison, we refer to \citet{bunge2013survey} and \citet{yee2015vgam}.

\begin{table}[h]
    \centering
    \scalebox{0.89}{\begin{tabular}{|l|c|c|c|c|c|c|}\hline
        {R} package & cont. covariate & variance formula & populn. type & param. & nonparam. & eff. \& DR\\\hline
        {Rcapture} 2007 & & & closed/open & \checkmark & &\\
        {RMark} 2008 & \checkmark & \checkmark & closed/open & \checkmark & &\\
        {CARE1} 2014 & & & closed & & \checkmark &\\
        {VGAM} 2015 & \checkmark & \checkmark & closed & \checkmark & &\\
        {drpop} & \checkmark & \checkmark & closed & \checkmark & \checkmark & \checkmark\\\hline
    \end{tabular}}
    \caption{This table lists some {R} packages for population size estimation. This list is not exhaustive. Our main focus is on the closed population set-up with discrete capture times. We have listed some properties like whether the package can incorporate individual level continuous covariate, has a closed form variance formula, population type it is applicable to, whether it can fit parametric/nonparametric model, and whether it is efficient and doubly robust.}    \label{tab:packagelist}
\end{table}

\subsection{Advantages of drpop}

The main goal of {drpop} is to improve estimation while using complex covariate information to model the heterogeneity. Unlike existing software, the methods in {drpop} are fully nonparametric, doubly robust, and optimally efficient under weak nonparametric conditions \citep{das2021doubly}.  {drpop} also lets the user apply their choice of flexible model(s) to capture the heterogeneity in the data. Moreover, it is applicable for data with any number of lists and works with arbitrary discrete or continuous covariates. 

In terms of usability, one of the attractions of {drpop} is that it comes with a lot of options for customization, starting from the model to the level of precision in the estimation. The user can select one or more model(s) for the covariates. The package comes with six in-built models, and is also capable of accepting user-provided model estimates. Further, {drpop} provides the user with the option to return a baseline estimator and an alternate targeted maximum likelihood estimator \citep{van2006targeted} in addition to the proposed doubly robust estimator. In the presence of categorical or numeric discrete covariates, one can also obtain estimates for sub-populations. Other than estimates, there is also an in-built function to simulate data to test models and a plot function for easy inference.

\subsection{Overview of paper}

In this paper, we present the package and some of its applications. Starting in section \ref{sec:setup}, we discuss the data structure for capture-recapture problems and introduce the necessary notations and the identification assumption. In section \ref{sec:method}, we briefly present the estimation method from \citet{das2021doubly} to obtain a doubly robust efficient estimator and the formula to obtain a confidence interval. Following this in section \ref{sec:implement}, we present some examples on how to use the {drpop} for different data types or problems and interpretation of the results. Section \ref{sec:perform} presents some error rates and performance comparison with some commonly used existing packages to motivate the use of {drpop}.

\section{Set-up}
\label{sec:setup}
In this section, we will discuss the data structure for the capture-recapture data we use. Depending on the approach, there are multiple ways to structure capture-recapture data. In the first subsection, we present our data structure and introduce some of the important notations. In the next subsection, we will discuss the identifiability assumption that the data must satisfy for valid estimates.

\subsection{Data structure}\label{sec:datastructure}

For a typical capture-recapture problem, the data is a collection of multiple lists. The lists contain information of the capture history of the observed/capture individuals/units. We use $K$ to denote the number of lists. We denote the unknown total population size by $n$ and the number of observed individuals by $N$. For observed individual $i$, $i\in \{1,\dots, N\}$, the capture history is a $K$-length vector of indicators $\bY_i = (Y_{i1}, \dots, Y_{iK})$. $Y_{ik}$ is 1 if individual $i$ is captured/observed in list $k$ and 0 otherwise. One individual can appear in multiple lists simultaneously, but an observed individual must appear in at least one of the lists i.e. $\bY_i \neq \bzero$.

In addition to capture profile i.e., lists, we consider the case where we also have covariate information for the observed individuals. We denote the covariate (or covariate vector) for individual $i$ by $\bX_i$, which can be used to model the individual-level heterogeneity. We thus  denote all data for individual $i$ by $\bZ_i = (\bY_i, \bX_i)$, and we assume $\bZ_i \sim \Pb$ independently.

The observed data size $N$ is a random draw from the binomial distribution $Binomial(n, \psi)$, where $\psi$ is the capture probability defined by
$$\psi\equiv\Pb(Y_1 \vee Y_2 \vee\dots\vee Y_K = 1)= \Pb(\bY \neq \bzero).$$
The capture probability $\psi$ is the probability of being observed in at least one of the $K$ lists. By the property of binomial distribution, any estimator for $\psi$ can be transformed to obtain an estimator for $n$ as follows
$$\widehat{n} = N/\widehat\psi.$$
However, since we only observe the individuals who satisfy $\bY \neq \bzero$, we cannot estimate $\Pb$, and hence, $\psi$ and $n$ directly. Instead, we can estimate the observed data distribution $\Qb$, where $\Qb$ at a point $\bz = (\by, \bx)$ is defined as
$$\Qb(\bY = \by,\, \bX = \bx) = \Pb(\bY = \by,\, \bX = \bx \ | \ \bY \neq \bzero) = \frac{\Pb(\bY = \by,\, \bX = \bx)\one(\by \neq \bzero)}{\psi}.$$

If we have $d \ (>1)$ dimensional covariates, then the data matrix is of dimension $N\times (K + d)$; each unique individual in its own row. In table \ref{tab:datastructure}, we present a typical capture-recapture data. The first $K$ columns denote the $K$ lists i.e., data source. The remaining $d$ columns contain the covariate information.

\begin{table}[ht]
\centering
\scalebox{1}{
\begin{tabular}{c|cccccccc}
observed individuals & list 1 & list 2 &\dots &list $K$ & \multicolumn{3}{c}{covariate(s)}\\\hline
1 & $Y_{11}$ & $Y_{12}$ & \dots & $Y_{1K}$ & $X_{11}$ & \dots &$X_{1d}$ \\
2 & $Y_{21}$ & $Y_{22}$ & \dots & $Y_{2K}$ & $X_{21}$ & \dots & $X_{2d}$ \\
\vdots &\vdots &\vdots&\vdots&\vdots&\vdots & \vdots & \vdots\\
N & $Y_{N1}$ & $Y_{N2}$ & \dots & $Y_{NK}$ & $X_{N1}$ & \dots & $X_{Nd}$ \\
\hline
\end{tabular}}
    \caption{A typical capture recapture data set from a population with $N$ observed individuals. The data is collected over $K$ sessions or using $K$ sources (lists). Each individual has one or more covariates ($d$ in this example).}
    \label{tab:datastructure}
\end{table}

\subsection{Identifiability}\label{sec:identifiability}

As discussed in the previous section, for capture-recapture data, we cannot directly estimate the unconstrained underlying distribution $\Pb$. Instead, we can estimate the observed data distribution $\Qb$. Further, to shift from $\Qb$ to $\Pb$, we need additional assumptions to ensure identifiability. In general, we assume some lack of dependence among the $K$ lists.

The simplest and oldest capture-recapture problems considered only $K = 2$ lists and had no covariates. One can assume that the two lists are independent i.e., $Y_1 \ind Y_2$ to ensure identifiability. This set-up has been used in \citet{petersen1896yearly}. However, the earliest known instance of this approach is by Graunt in the 1600s \citep{hald2003history} followed by Laplace \citep{goudie2007captures}. It has been further extended to the more than three list case by \citet{Darroch1958indeplists} and \citet{schnabel1938estimation}. There have been other modifications to this approach over the years \citep{jolly1983problem, seber1982estimation, bailey1952improvements}. For more discussion, we refer to \citet{krebs2014ecological}. 

Note that it is important that the lists are not completely dependent, i.e., they must have some overlap and they must not be identical, to say the least. Both these cases are uninformative of the unobserved population, and contain the same amount information as the case when we observe only one list. Thus, to ensure identifiability of the total population size, we need some lack of dependence assumption among the lists.

\citet{das2021doubly} assumes that two lists out of the $K$ lists are collected independently conditioned on the covariate(s). Without loss of generality one can assume that lists 1 and 2 are conditionally independent. One can always reorder the columns to have the two conditionally independent list pair at position 1 and 2. This assumption has been used very often in past work \citep{tilling1999capture, sekar1949method, alho1993estimating, huggins1989statistical, chao1987estimating, pledger2000unified, burnham1979robust, pollock1990statistical, huggins2007non}.

\begin{assumption}\label{as:1}
$\Pb( Y_1 = 1 \mid \bX=\bx, Y_2=1) = \Pb(Y_1 = 1 \mid \bX=\bx, Y_2=0)$, where $Y_k$ denotes the capture indicator variable for list $k$ for $k = 1,\dots, K$.
\end{assumption}

This conditional independence assumption is more flexible compared to the conventional marginal independence assumption and accommodates a wide scenario of data collection procedure including the case when the lists have some kind of interaction. For example, when one is collecting data on documented patients at say two different hospitals. Then patients who have already been observed at hospital 1 might have a low probability of being observed again at hospital 2 and vice versa. Hence, the lists of the hospitals are not behaving independently. Now, if we have access to say the location information of the patients, we can describe the behavior of the patients conditioned on that i.e., patients are more likely to visit the hospitals nearer to them. Hence, conditioning on the location, one can assume independence between the two lists.

Another very common identifiability assumption in the capture-recapture literature is the log-linear model introduced by \citet{Fienberg1972loglinear}. There identifiability is ensured by assuming that the highest order interaction term among all the lists is zero. We refer to \citet{tilling1999capture, huggins2011review} for more discussion on the differences between identifying assumptions like conditional independence versus log-linear model-based dependence.
In particular we refer to  \citet{you2021estimation}, who give discussion and present methods in a general identification framework without covariates.

In the presence of covariates, we can define the conditional capture probability of an individual by $\gamma(\bx) = \Pb(\bY \neq \bzero \ | \ \bX = \bx)$. It is known \citep[e.g., as in][]{tilling1999capture} that under Assumption \ref{as:1} the capture probability $\psi$ can be identified from the biased observed data distribution $\Qb$. Specifically, let
\begin{align*}
q_{1}(\bx) &= \Qb(Y_1 = 1  \mid \bX=\bx) \\
q_{2}(\bx) &= \Qb(Y_2 = 1  \mid \bX=\bx) \\
q_{12}(\bx) &= \Qb(Y_1 = 1, Y_2 = 1 \mid \bX=\bx) 
\end{align*}
denote the observational probability (under $\Qb$) of appearing on list 1, 2, and both, respectively. These probabilities will be referred to as the $q$-probabilities at various points throughout. They are also called the nuisance functions or nuisance parameters in this problem set-up and, are crucial in the estimation process.

Now, under Assumption \ref{as:1}, we can define the conditional capture probability and the marginal capture probability as follows
\begin{align}
\gamma(\bx) &\equiv {\Pb(\bY \neq \bzero \mid \bX=\bx)} = \frac{q_{12}(\bx)}{q_{1}(\bx) q_{2}(\bx)} \\
\psi &\equiv  {\Pb(\bY \neq \bzero)} = \left\{ \int \gamma(\bx)^{-1} \ d\Qb(\bx)  \right\}^{-1}.
\end{align}

Using the expression on the right hand side above, we can directly estimate the capture probability $\psi$ and hence, the total population size by using $N/\psi$ from the observed data. We present the baseline and the proposed method in the following section.

\section{Methodology}\label{sec:method}

In this section, we discuss a simple plug-in estimator  and some of its disadvantages. Following that we discuss our new proposed method in \citet{das2021doubly} and the ways in which it improves upon the plug-in. Under assumption \ref{as:1}, \citet{das2021doubly} presents two different estimators for $\psi$ and $n$:  (i) a doubly robust (DR) estimator and (ii) a targeted maximum likelihood estimator (TMLE). 

The simplest estimator we can obtain from the expression of $\psi$ in the previous section is based on the plug-in principle, i.e., taking the identifying expression and constructing an estimator by replacing unknown quantities with estimates. The plug-in estimators for the capture-probability $\psi$ and the total population size $n$ are therefore 
$$\widehat\psi_{PI} = \left\{\sum_{i=1}^N \frac{\widehat{q}_{1}(\bx_i) \widehat{q}_{2}(\bx_i)}{\widehat{q}_{12}(\bx_i)}\right\}^{-1} \ \text{ and } \ \widehat{n}_{PI} = \frac{N}{\widehat\psi_{PI}},$$
where $\widehat{q}_j$ is the estimated probability value of $q_j$ for $j \in \{1,\, 2,\, 12\}$ and $\bx_i$ is the covariate value for the observed individual $i$. In principle, the $\widehat{q}_j$ can be estimated with any parametric (logistic, multinomial logistic) or nonparametric (random forest, gradient boosting) models, though the performance of the plug-in can vary greatly depending on what kind of model is used.

In particular,  plug-in estimators typically inherit  mean squared errors of the same order as their nuisance parameter estimates $\widehat{q}_j$. This means that when using a plug-in the problem of estimating the one-dimensional capture probability/population size is often made as difficult as estimating the $d$-dimensional $q$-probabilities. If one has correct parametric models for these probabilities, this is of little concern, but correct parametric models are hard to come by in practice. When using more flexible methods like random forests or gradient boosting, one would inherit the larger mean squared errors necessarily obtained in nonparametric regression problems.  

Beyond the issue of plug-ins having potentially large mean squared errors, in general they also do not come with closed-form variance formulas, which is important for constructing confidence intervals. The bootstrap can be used when parametric models are used to estimate the $q$-probabilities \citet{tilling1999capture}, but in general the bootstrap fails when more flexible methods are used (e.g., ensembles of high-dimensional regressions).

The proposed doubly robust estimator in \citet{das2021doubly} uses elements from semiparametric theory \citep{tsiatis2006semiparametric, bickel1988estimating, kennedy2016semiparametric, van2006targeted, van2002semiparametric} to tackle some of the deficiencies of the plug-in estimator discussed above. We discuss more about these properties in the following section.

\subsection{Proposed Estimators}

\citet{das2021doubly} proposed a doubly robust estimator using semiparametric theory and influence functions. More details on general  efficiency theory can be found in \citet{bickel1993efficient}, \citet{van2002part}, and \citet{van2003unified}; reviews can be found in \citet{tsiatis2006semiparametric} and \citet{kennedy2016semiparametric} among others.

\citet{das2021doubly} showed that the (uncentered) efficient influence function of the capture probability $\psi$ is given by
$$\phi_i = \frac{1}{\gamma(\bX_i)}\left\{ \frac{Y_{1i}}{q_1(\bX_i)} + \frac{Y_{2i}}{q_2(\bX_i)} - \frac{Y_{1i}Y_{2i}}{q_{12}(\bX_i)}\right\},$$
where $\gamma(\bX_i) = \frac{q_{12}(\bX_i)}{q_1(\bX_i)q_2(\bX_i)}$ is the conditional capture probability of observation $i$. The efficient influence function is crucial since (i) its variance acts as a minimax lower bound in nonparametric models \citep{van2002part},  and (ii) it can be used to construct efficient estimators that attain the minimax lower bound. Since the expected value of the efficient influence function is the inverse capture probability $\psi^{-1}$, \citet{das2021doubly} proposed the following doubly robust estimators for the capture probability and the total population size $n$
$$\widehat\psi_{DR} = \left(\frac{1}{N} \sum_{i=1}^N \widehat\phi_i\right)^{-1} \ \text{ and } \ \widehat{n}_{DR} = \frac{N}{\widehat{\psi}_{DR}},$$
where $\widehat\phi_i$ is obtained by substituting the estimates of the $q$-probabilities into $\phi_i$. This estimator has some very favorable properties such as: (i) $1/n$-rate mean squared errors, even in flexible non-parametric models, (ii) double robustness, (iii) local asymptotic minimaxity, and (iv) asymptotic normality with finite-sample guarantees. We briefly discuss these properties in this paper, and for more details refer to \citet{das2021doubly}.

As a consequence of efficiency theory, the error in estimation using the proposed estimator is of the order of $1/\sqrt{n}$ even when all three nuisance parameters are estimated flexibly. The formal result states that for any sample size $N$ and error tolerance $\delta>0$,
$ | (\widehat\psi^{-1}_{dr} - \psi^{-1}) - \Qn \phi | \leq \delta $
with probability at least
$ 1 - \left( \frac{1}{\delta^2} \right) \E\left( \widehat{R}_2^2  + \frac{ \| \widehat\phi - \phi \|^2 }{N} \right) $. $\widehat{R}_2$ is a second-order error term given by
\begin{align*}
\widehat{R}_2 &= \int \frac{1}{\widehat{q}_{12}} \left\{ \Big( q_1 - \widehat{q}_1 \Big) \Big( \widehat{q}_2 - q_2 \Big) + \Big( q_{12} - \widehat{q}_{12} \Big) \left( \frac{1}{\gamma} - \frac{1}{\widehat\gamma} \right) \right\} \ d\Qb \\
& \leq \left( \frac{1}{\epsilon} \right) \| \widehat{q}_1 - q_1 \| \| \widehat{q}_2 - q_2 \| + \left( \frac{1}{\epsilon^3} \right) \| \widehat{q}_{12} - q_{12} \| \| \widehat\gamma - \gamma \| 
\end{align*}
The bound on $\widehat{R}_2$ holds as long as $(q_{12} \wedge \widehat{q}_{12}) \geq \epsilon$. If all the nuisance parameters are estimated with error $\sim n^{-1/4}$, i.e., a rate commonly found in nonparametric regression problems, then the error is still bounded above by $C/\sqrt{n}$ for some constant $C$ with probability converging to 1 as $n$ and therefore $N$ increases. The plugin estimator however, does not possess this property and in general inherits the slower rate (e.g., $n^{-1/4}$) from the nonparametric estimation of the $q$-probabilities. 

Another important property of the proposed estimator is the double robustness property presented in corollary 2 in 
\citet{das2021doubly}. This result follows directly from the formula of the second order error term above $\widehat{R}_2$. This result states that if two out of the four quantities $\gamma,\, q_1,\, q_2$, and $q_{12}$ have small estimation error, then $\widehat\psi_{DR}$ and hence, also $\widehat{n}_{DR}$ will have small estimation error. More specifically, we need one of $q_{12}$ and $\gamma$ to be estimated with small error and, one of $q_1$ and $q_2$ to be estimated with small error. This property is useful when one of the two lists is difficult to estimate or is a complex function of the covariates. More details can be found in  \citet{das2021doubly}.

Since the proposed estimator $\widehat\psi_{DR}$ is a sample average of the estimated efficient influence functions, it has variance nearly equal to the variance of the estimated efficient influence function divided by $N$,  i.e.,
$$\var(\widehat\psi_{DR}^{-1}) = \frac{\var(\widehat\phi)}{N}.$$
If $\sigma^2$ denotes the population variance of $\phi$, then one can estimate $
var(\widehat\psi_{DR}^{-1})$ by $\widehat\sigma^2/N$ where $\widehat\sigma$ denotes the estimator of $\sigma$. The variance of the efficient influence function divided by $N$, i.e., $\sigma^2/N$, acts as a minimax lower bound in the sense that it is the lowest possible mean squared error any estimator can achieve in a local neighbourhood. The mean squared error of the proposed estimator is close to this bound for a large sample size, e.g., when the nuisance parameters are estimated with errors converging to zero. \citet{das2021doubly} further presented finite sample analogs of the usual asymptotic minimax arguments and error bounds, including finite-sample distance from a Gaussian distribution.

All the properties we discussed for the capture probability estimate also apply to the total population size estimate. In population size estimation problems, the main interest is often in a confidence interval for $n$. In the next section, we discuss the properties of the estimated confidence interval.

\subsubsection{Confidence interval estimation}

One of the main motivations behind using the proposed estimator is that it has a well defined variance formula, as discussed in the previous section. One can estimate $\var(\widehat{\psi}^{-1}_{DR})$ by using the unbiased sample variance of $\widehat\phi$ scaled by $N$. This can be used to obtain the variance estimator for the estimated total population size $\widehat{n}_{DR}$ which is given by
$$\widehat\var(\widehat{n}_{DR}) = N\widehat\var(\widehat\phi) + \frac{N(1 - \widehat\psi_{DR})}{\widehat\psi_{DR}^2},$$
where $\widehat\var(\widehat\phi)$ is the unbiased variance estimate of $\widehat\phi$. For the derivation, we refer to \citet{das2021doubly}. This variance formula for the estimated total population size can be applied more generally to any estimator that can be approximated by a sample average. The estimated $(1-\alpha)\times 100\%$ confidence interval is
$$\widehat{CI_n} = \widehat{n} \pm z_{\alpha/2} \sqrt{\widehat\var(\widehat{n}_{DR})}.$$
The finite sample validity/coverage error for this interval is presented in \citet{das2021doubly}. In particular they show the coverage error is bounded above as 
$$ \left|\Pb\left( \widehat{\text{CI}_n} \ni n \right) - (1-\alpha)\right| \ \lesssim \ {n^{(1-4\beta)/2}} + \frac{1}{\sqrt{n}} , $$
if the nuisance estimators have mean squared errors of order $O(n^{-2\beta})$. Hence, if $\beta>1/4$, then for any $\epsilon > 0$, there exists an $N_\epsilon$, such that the coverage error is less than $\epsilon$ for any $N > N_\epsilon$.

The availability of a closed form formula for the variance and hence, the confidence interval allows for simple inference. Moreover, this also can be used to study the effect of the constituent elements on the variance of the estimate. This eliminates the need to use methods like bootstrap, for example, which can be computationally intensive or not guaranteed to provide valid coverage.

\citet{das2021doubly} also presented an alternate targeted maximum likelihood estimator $\widehat\psi_{TMLE}$ and the associated $\widehat{n}_{TMLE} = N/\widehat\psi_{TMLE}$. This estimator has the same properties as $\widehat\psi_{DR}$, but the method of calculation uses clever covariates in the targeted maximum likelihood algorithm \citep{van2006targeted, van2011targeted}. This estimator does not have a closed form expression. For simplicity, we focus on the original doubly robust estimator in this paper.

\section{Implementation using drpop}\label{sec:implement}

In this section, we illustrate the various functions available in {drpop} and their implementation in detail. The main goal of {drpop} is to easily evaluate a doubly robust efficient estimate of the total population size and an associated confidence interval from any capture-recapture data with covariate information. The package is capable of handling high-dimensional and/or complex covariates, both discrete and continuous. It also contains some additional functions that aid in method design, model testing, and inference.

\begin{figure}[t]
    \centering
    \includegraphics[width = 12.9cm, height = 6.6cm]{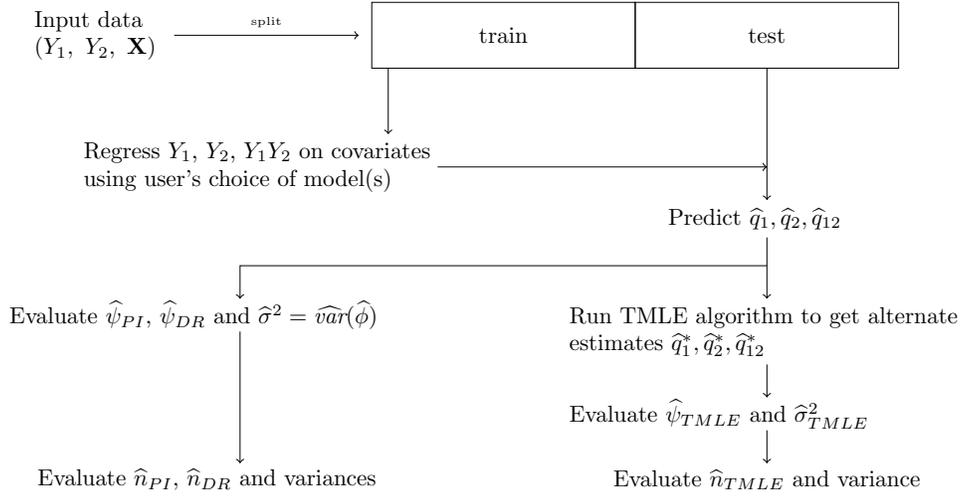}
    \caption{The above figure depicts the estimation procedure followed by the estimation function in the package. For simplicity, we show only two lists, and only one train and one test sample. The functions in the package however, uses cross-fitting to utilize the whole observed data.}
    \label{fig:estim_flowchart}
\end{figure}

Before diving into the implementation, we discuss the estimation process for a given dataset. In the previous section, we discussed three possible estimators: the plug-in (PI), the proposed doubly robust (DR) and the targeted maximum likelihood estimator (TMLE). {drpop} has the option to return all three of these estimators, though the default is just to return the doubly robust estimator. To illustrate the steps in the estimation process, we present a flow chart in Figure \ref{fig:estim_flowchart} that evaluates the estimates for the capture probability $\psi$ and the total population size $n$ for a capture-recapture dataset with two lists. For the case of more than two lists ($K>2$), {drpop} returns the estimates for every possible list-pair unless specified otherwise. Moreover, {drpop} uses cross-fitting to achieve complete efficiency \citep{zheng2010asymptotic, robins2008higher, chetverikov2021cross}. But, for simplicity, we only present a simple sample-splitting in the flow chart.

Following is the list of functions available in the package along with their brief descriptions.
\begin{enumerate}
    \item \code{simuldata}: Generate two or three list toy data with desired features
    \item \code{informat}: Check if data is in format
    \item \code{reformat}: Reorder columns to put data in format
    \item \code{qhat\_logit}, \code{qhat\_mlogit}, \code{qhat\_gam}, \code{qhat\_ranger}, \code{qhat\_sl}, \code{qhat\_rangerlogit}: Estimate nuisance parameters $q_1, q_2, q_{12}$
    \item \code{tmle}: Obtain targeted maximum likelihood estimates of nuisance parameters
    \item \code{popsize}: Estimate population size from raw data or with user provided nuisance estimates
    \item \code{popsize\_cond}: Estimate population size from raw data conditional on a discrete covariate
    \item \code{plotci}: Plot the results of \code{popsize}, or \code{popsize\_cond}.
\end{enumerate}
For a given dataset, one only needs to call either \code{popsize} or \code{popsize\_cond} to get the estimates of the capture probabilities, total population size, and the confidence intervals.

In this section we briefly describe some data types one can come across and the interpretation of the results. To illustrate the use, we will use toy data examples. A typical dataset in the capture-recapture format has at least two binary columns (corresponding to two or more lists) indicating list-wise capture profiles and one or more covariate column(s). Each observed or captured individual has their own row.

\subsection{Choice of models for nuisance parameters}\label{sec:modelnuis}

The estimation of the population size and the capture probability requires modelling the capture profiles conditional on the covariates. {drpop} provides six modelling choices listed as follows.
\begin{enumerate}
    \item \code{logit}: Fits logistic regression using {R} function \code{glm}.
    \item \code{mlogit}: Fits multinomial logistic regression using {R} function \code{multinom} in package \code{nnet}.
    \item \code{gam}: Fits simple generalized additive model from the {R} package \code{gam}.
    \item \code{ranger}: Fits random forest model from the {R} package \code{ranger}. Suitable for high dimensional covariates.
    \item \code{rangerlogit}: Fits an ensemble of random forest and logistic model.
    \item \code{sl}: Fits different SuperLearner algorithm from the library provided by the user from the {R} package \code{SuperLearner}. Returns estimates using a combination of the fitted models. The user can specify the library of models via \code{sl.lib}.
\end{enumerate}

The computation time varies based on the above models. The parametric models \code{logit} and \code{mlogit} are generally the fastest. However, they can lack flexibility, making resulting  estimates  biased if the nuisance parameters are more complex functions of the covariates. \code{gam} is slightly slower than the parametric models, but is still comparably fast enough for practical purposes. The flexible nonparametric models \code{ranger} and \code{rangerlogit} can be slower to run than these previous models. However, being flexible, these methods can accommodate more complex nuisance functions. \code{rangerlogit} is the default model in {drpop} and the performance statistics are presented in section \ref{sec:perform}. \code{sl} is the slowest depending on the models passed into \code{sl.lib}. This is because it aggregates multiple models, returning the best estimator combining the individual models using cross-validation. {drpop} provides the user with the option to parallelize \code{sl} using \code{snowSuperLearner} from the {R} package \code{SuperLearner}, which is supported on all three of Windows, MacOS and Linux.

For simplicity, we apply some of these models on a toy dataset, \code{listdata} as shown below. The true population size is 2000 and there are $N=1610$ rows in the data. The columns \code{y1}, \code{y2} and \code{x1} show list 1 captures, list 2 captures and a continuous covariate. The empirical capture probability is approximately 0.85. 

\begin{Verbatim}[fontsize = \normalsize]
> head(listdata, 3)
  y1 y2       x1
1  1  1 2.159287
2  0  1 2.654734
3  1  1 5.338062

\end{Verbatim}

The function \code{popsize} returns the estimates via \code{nuis} for the observed data probabilities $q_1$, $q_2$ and $q_{12}$ which are often called the nuisance estimates. It also returns the fold assignment for each row. For simplicity, we use two folds and plot the estimated nuisance parameters.

\begin{Verbatim}[fontsize = \normalsize]
> qhat = popsize(data = listdata, funcname = c("rangerlogit", "logit",
                 "gam", "mlogit", "sl"), nfolds = 2)
\end{Verbatim}

\begin{figure}[t]
    \centering
    \includegraphics[width = \textwidth]{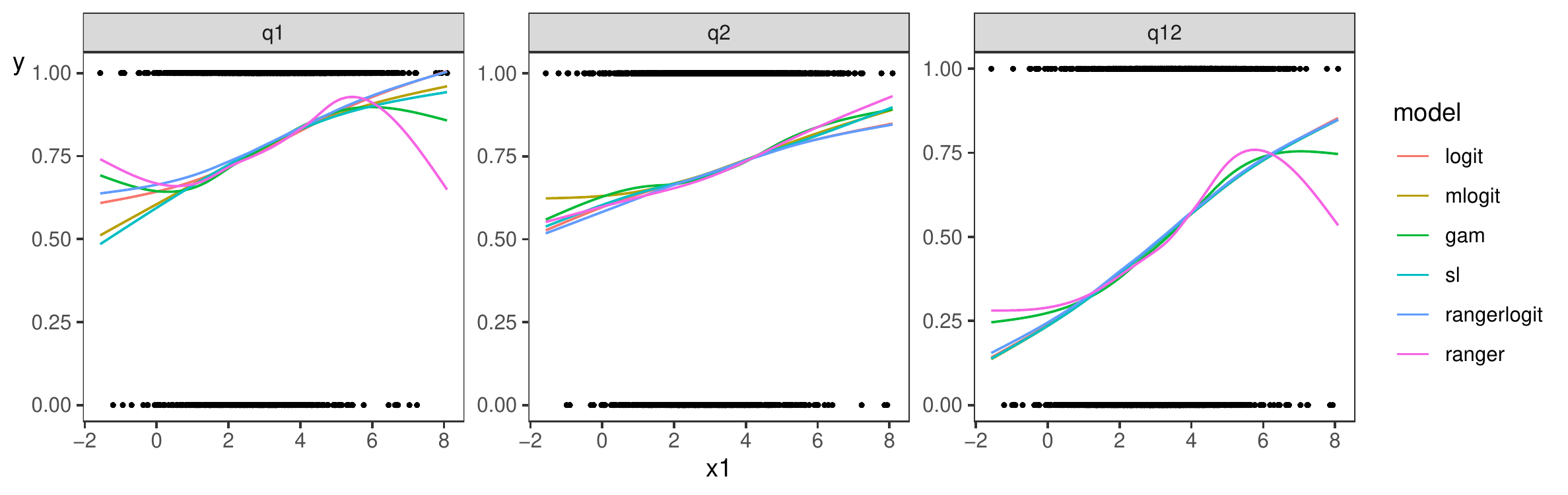}
    \caption{The plot shows the smoothed estimated \code{q1}, \code{q2} and \code{q12} for five different models against the scalar covariate \code{x1}. The points at 0 and 1 show the capture profiles of the individuals i.e., $Y_1$, $Y_2$ and $Y_1Y_2$ respectively.}
    \label{fig:qhat}
\end{figure}

The dataframe \code{qhat\$nuis} contains the estimates for $q_1$, $q_2$ and $q_{12}$ for each model supplied by the user for each row of the data. \code{qhat\$idfold} shows the fold assigned to each row. Figure \ref{fig:qhat} shows the estimated probabilities along with the capture profiles of list 1, list 2 and the two lists simultaneously. One also has the option of using models outside the {drpop} package and obtain estimates which we will present later in section \ref{sec:estimatenuis}. Next, we illustrate some examples of application of the package starting from the simplest case.

To ensure that the estimator is valid, we required that all the nuisance parameter estimates, which are probabilities, are bounded away from zero. The default bound is 0.005. One can change this using the argument \code{margin} in \code{popsize} or \code{popsize\_cond}.

\subsection{Two-list case with covariates}\label{sec:2list}

The simplest capture-recapture data has two lists with one or more covariates. We present the toy data, \code{listdata} with true population size 5000 and two continuous covariates.

\begin{Verbatim}[fontsize = \normalsize]
> head(listdata, 3)
  y1 y2       x1       x2
1  1  1 5.342829 0.4682059
3  1  0 3.700239 2.0279143
4  1  1 4.279882 3.3915513
> result = popsize(data = listdata, funcname = c("logit", "gam", "mlogit", "sl"))

\end{Verbatim}

To obtain the total population size estimate, we call the function \code{popsize}. This function accepts the data frame \code{listdata} as \code{data} and list of model names, \code{funcname} which are to be used to estimate the nuisance parameters ($q_1$, $q_2$, $q_{12}$). \code{popsize} returns a list of objects which include the estimated population size, estimated capture probability, estimated variance and the 95\% confidence intervals. Above we print only the estimated capture probabilities \code{psi}, estimated population sizes \code{n}, estimated $\sigma$ \code{sigma}, estimate standard deviation of $\widehat{n}$ \code{sigman} and the 95\% confidence intervals \code{cin.l, cin.u} for the total population size. The columns \code{listpair, model} and \code{method} indicate the list pairs (lists 1 and 2 in this case), model used to estimate heterogeneity from covariates, and the formula for estimation of the target parameters $\psi$ and $n$ respectively.

\begin{remark}
Setting arguments \code{PLUGIN} and \code{TMLE} to \code{FALSE} will return only the \code{DR} (proposed doubly robust) estimates. We also plot the confidence intervals using the \code{plotci} function.
\end{remark}

\begin{Verbatim}[fontsize = \normalsize]
> result = popsize(data = listdata, funcname = c("gam", "logit",
           "mlogit", "sl"), PLUGIN = TRUE, TMLE = TRUE)
> print(result)
   listpair  model method   psi sigma    n  sigman cin.l cin.u
1       1,2    gam     DR 0.910 0.440 4978  37.032  4905  5051
2       1,2    gam     PI 0.917 0.440 4941  36.433  4870  5012
3       1,2    gam   TMLE 0.912 0.595 4968  45.649  4878  5057
4       1,2  logit     DR 0.910 0.478 4978  39.073  4901  5054
5       1,2  logit     PI 0.918 0.478 4936  38.423  4860  5011
6       1,2  logit   TMLE 0.900 1.670 5034 114.852  4809  5259
7       1,2 mlogit     DR 0.910 0.498 4978  40.184  4899  5056
8       1,2 mlogit     PI 0.908 0.498 4986  40.311  4907  5065
9       1,2 mlogit   TMLE 0.897 1.875 5052 128.443  4800  5304
10      1,2     sl     DR 0.910 0.452 4979  37.689  4905  5053
11      1,2     sl     PI 0.917 0.452 4938  37.034  4865  5010
12      1,2     sl   TMLE 0.896 1.954 5054 133.735  4792  5316
> plotci(result)

\end{Verbatim}

\begin{remark}
Since the plug-in estimator has no known variance formula, we use the same variance formula as the proposed estimator for the calculation of the variance of the plug-in estimators.
\end{remark}

\begin{figure}[t]
    \centering
    \includegraphics[width = \textwidth]{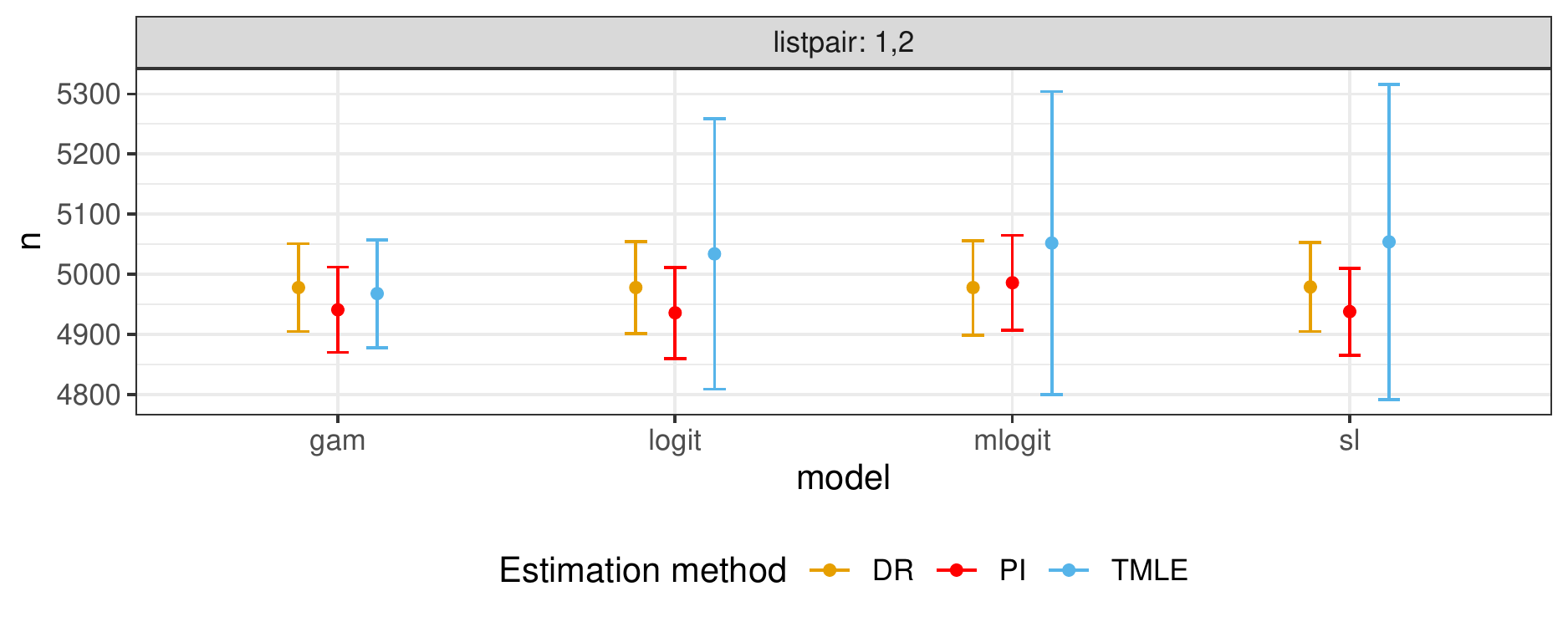}
    \caption{The above plot shows the estimated confidence interval for $n$ for different models. The true population size is 5000. The term list-pair specifies the two lists used for the estimation.}
    \label{fig:cin_2list}
\end{figure}

\subsection{Two-list case with conditional estimates}\label{sec:2listcat}

When one has a discrete or categorical covariate in addition to other covariates, it is often of interest to estimate the total population size conditioned on that categorical covariate, i.e., for sub-populations. For example, suppose one has a population of patients in a city and their age, demographic information, and ethnicity as the covariates. Then it can be of interest to obtain the estimated population size for the different ethnicities separately.

We again use a simulated toy dataset to illustrate the implementation. The data has three continuous covariates (\code{x1, x2, x3}) and one categorical covariate column called \code{catcov}. \code{catcov} takes three possible values `a', `b', `c' with equal probability. Total population size is 6000 and each of `a', `b' and `c' appear roughly 2000 times in the whole population. We present the first three rows below.

\begin{Verbatim}[fontsize = \normalsize]
> head(listdata, 3)
  y1 y2       x1       x2        x3 catcov
1  1  1 2.159287 5.897364 3.4173336      b
2  1  0 2.654734 2.075288 0.5961934      a
3  1  0 5.338062 2.156149 2.5186507      c

\end{Verbatim}

The interest here is to obtain population size estimates conditioned on the categorical variable \code{catcov}, i.e., for sub-populations with \code{catcov} value `a', `b' and `c' separately. The function \code{popsize\_cond} is similar to the function \code{popsize} but returns the result separately for each level of the categorical variable. We specify the categorical covariate to be used for conditioning using the argument \code{condvar}. To obtain an overall estimate one can use \code{popsize} as in the previous example.
\begin{Verbatim}[fontsize = \normalsize]
> result = popsize_cond(data = listdata, condvar = 'catcov',
           funcname = c("mlogit", "gam"), PLUGIN = TRUE, TMLE = TRUE)
> print(result)
   listpair  model method   psi  sigma    n  sigman cin.l cin.u condvar
        1,2 mlogit     DR 0.560  4.821 3040 204.818  2639  3442       b
        1,2 mlogit     PI 0.575  4.821 2960 204.323  2560  3361       b
        1,2 mlogit   TMLE 0.541  7.050 3147 295.398  2568  3726       b
        1,2     sl     DR 0.627  5.501 2715 230.476  2263  3167       b
        1,2     sl     PI 0.606  5.501 2808 230.926  2355  3260       b
        1,2     sl   TMLE 0.637  3.296 2670 141.464  2393  2948       b
        1,2 mlogit     DR 0.596  4.683 3306 213.115  2888  3724       a
        1,2 mlogit     PI 0.590  4.683 3338 213.290  2920  3756       a
        1,2 mlogit   TMLE 0.630  3.401 3126 156.890  2818  3433       a
        1,2     sl     DR 0.612  3.731 3216 171.610  2880  3553       a
        1,2     sl     PI 0.630  3.731 3125 171.019  2790  3460       a
        1,2     sl   TMLE 0.594  5.777 3313 260.696  2802  3824       a
        1,2 mlogit     DR 0.533  7.068 3082 291.147  2511  3652       c
        1,2 mlogit     PI 0.558  7.068 2946 290.526  2377  3516       c
        1,2 mlogit   TMLE 0.489 10.859 3359 444.129  2488  4229       c
        1,2     sl     DR 0.524  6.486 3138 268.283  2612  3664       c
        1,2     sl     PI 0.585  6.486 2807 266.666  2285  3330       c
        1,2     sl   TMLE 0.476 11.535 3453 471.596  2529  4377       c
> plotci(result)

\end{Verbatim}

The result of \code{popsize\_cond} is in a similar format to \code{popsize}, but it specifies the level of the categorical covariate i.e., the sub-population in a separate column.
\begin{figure}[t]
    \centering
    \includegraphics[width = \textwidth]{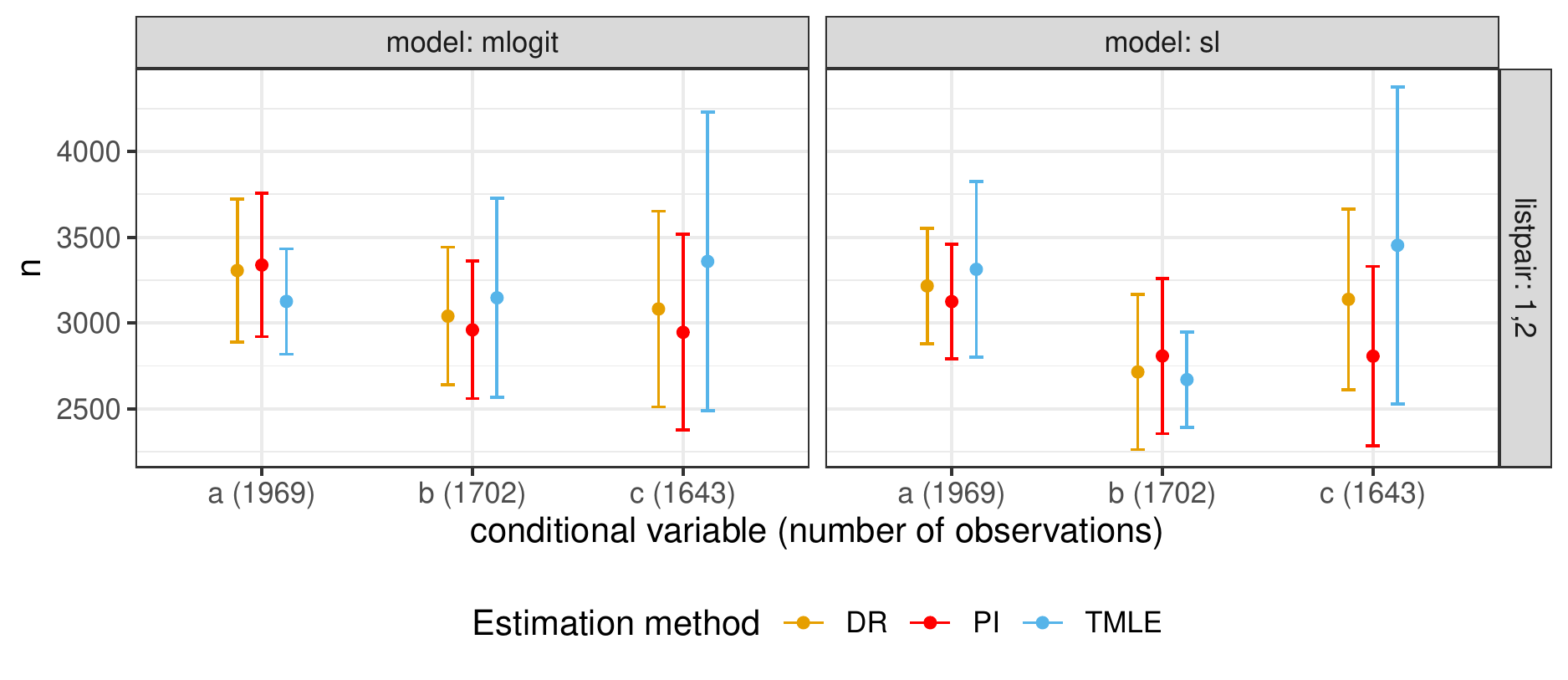}
    \caption{The above figure shows the confidence interval for $n$ for three sub-populations and models gam, logit and sl. The sub-populations are obtained from the original population using the values of the \code{catcov} covariate.}
    \label{fig:cin_cond}
\end{figure}

\subsection{Three or more lists}\label{sec:3list}

The approach used by {drpop} assumes that there are two lists which are known to be conditionally independent. However, capture-recapture datasets can often consist of more than two lists. If the analyzer knows the list-pair that is conditionally independent, they can use the functions \code{popsize} and \code{popsize\_cond} by removing the remaining list columns or by specifying the two list columns to be used for estimation. However, when the analyzer is not aware of the list-pair, the entire dataset can be passed into the estimation functions. {drpop} returns an estimate for each possible list-pair.

The toy dataset has three list columns as shown below. Now, since we pass more than two list columns into the functions, the output will have the result for the different list-pairs \code{(1,2), (1,3)} and \code{(2,3)}.

\begin{Verbatim}[fontsize = \normalsize]
> head(listdata,3)
  y1 y2 y3       x1       x2        x3       x4
1  0  0  1 1.189401 6.737728 0.8531169 1.508898
2  1  0  1 3.416144 3.079832 3.1891693 4.082209
3  1  0  0 4.626662 3.684374 3.6552886 2.694606

\end{Verbatim}
For more than two lists, we need to specify the number of list columns using \code{K} in \code{popsize}. For simplicity, we evaluate only the doubly robust estimators in this example and exclude the TMLE and the plug-in estimates. The \code{listpair} column in the result below specifies the list-pair used for the estimation. For example, assuming $Y_1\ind Y_2$ under the \code{rangerlogit}, we get $\widehat{n}_{DR} = 29,711$, and assuming $Y_1\ind Y_3$ under the \code{rangerlogit} model, we get $\widehat{n}_{DR} = 30,423$.
\begin{Verbatim}[fontsize = \normalsize]
> result = popsize(data = listdata, K = 3, funcname = c("mlogit", "gam",
                   "rangerlogit"), nfolds = 2)
> result
   listpair       model method   psi sigma     n  sigman cin.l cin.u
1       1,2         gam     DR 0.872 0.983 29752 171.595 29416 30089
4       1,2      mlogit     DR 0.873 1.497 29723 249.911 29233 30213
7       1,2 rangerlogit     DR 0.874 1.453 29711 243.040 29235 30187
10      1,3         gam     DR 0.860 1.565 30192 261.799 29679 30705
13      1,3      mlogit     DR 0.851 2.271 30502 373.116 29771 31233
16      1,3 rangerlogit     DR 0.853 2.137 30423 351.831 29734 31113
19      2,3         gam     DR 0.859 2.467 30236 403.749 29445 31027
22      2,3      mlogit     DR 0.859 3.452 30234 560.642 29135 31333
25      2,3 rangerlogit     DR 0.853 2.871 30449 468.210 29531 31367
> plotci(result)

\end{Verbatim}

\begin{figure}[t]
    \centering
    \includegraphics[width = \textwidth]{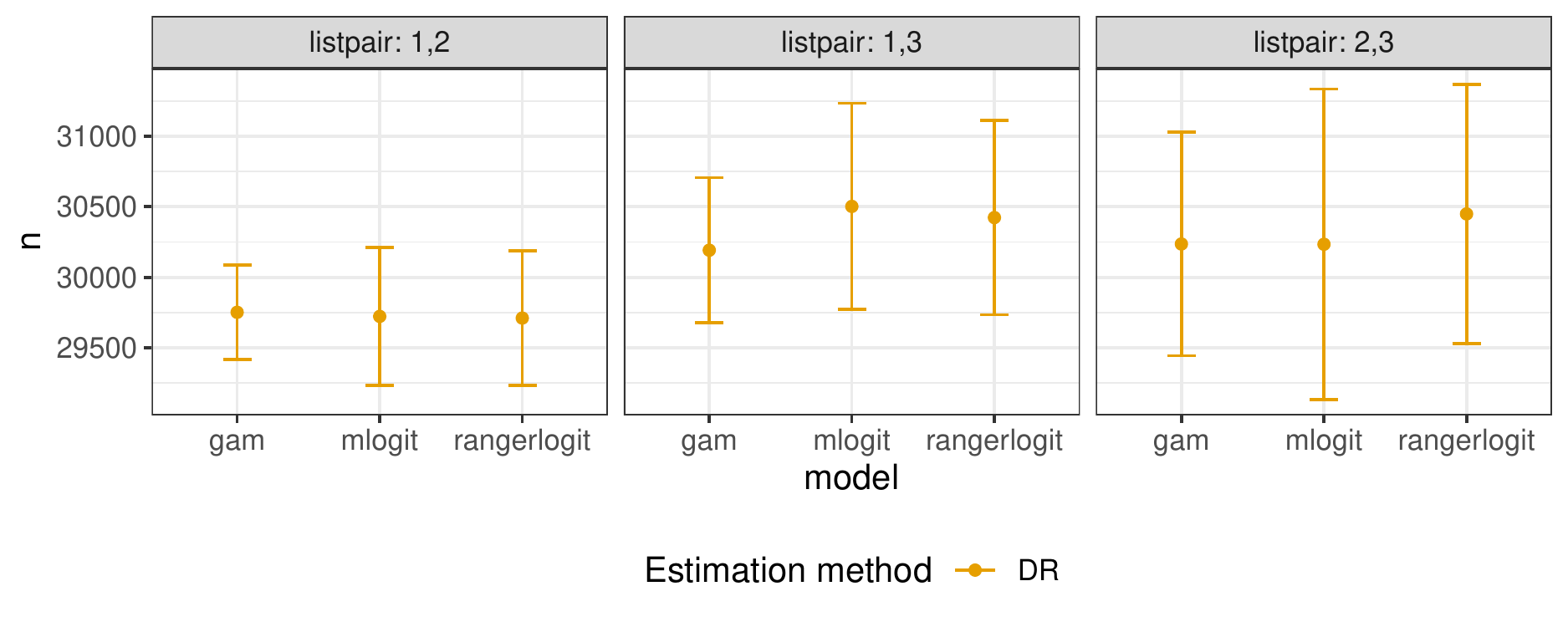}
    \caption{The above plot shows the estimated confidence interval for $n$ for three different possible list-pairs under different models. The result for list-pair (1,2) produces narrower intervals closer to the true value 30,000.}
    \label{fig:cin_3list}
\end{figure}

The plot function in the package shows the estimated confidence interval for $n$ in Figure \ref{fig:cin_3list}. We note that confidence intervals are relatively shorter for list-pair (1,2) and (1,3). The reason being that the overlap between the lists is larger for (1,2) and (1,3) compared to (2,3). As already discussed previously in section \ref{sec:setup}, the overlap between the conditionally independent lists must be bounded away from 0 and $N$ for better estimation.

If the analyzer is aware of the list pair, then the dataset can be passed into the estimation functions by removing all other list columns or by specifying the list pair. Suppose that the two conditionally independent list columns are \code{y1} and \code{y2}. Then the user can either remove column \code{y3} and pass the data into \code{popsize}, or he can specify the pair. We illustrate both these approaches below.

\begin{Verbatim}[fontsize = \normalsize]
> result = popsize(data = subset(listdata, select = -c(y3)))
> result = popsize(data = listdata, j = 1, k = 2, K = 3)

\end{Verbatim}

\subsection{Estimation with user provided nuisance estimates}\label{sec:estimatenuis}

The main purpose of this example is to illustrate how to pass nuisance parameter estimates into \code{popsize}. This is useful when the user has some background information that suggests modelling the heterogeneity differently than what is available in the package. For simplicity, we illustrate this by passing the nuisance parameter estimates, \code{nuis} from the output of \code{popsize} with the default model \code{rangerlogit}. The toy dataset used has total population size 5000 with two continuous covariates. We show the first few rows of the estimated nuisance parameters in \code{estim\$nuis}. The columns specify the model name (\code{rangerlogit} in this case) and also the $q$-probabilities. For more than one model, \code{estim\$nuis} will contain additional columns in the same format. \code{estim\$idfold} specifies the fold assignment for each row. Rows 1 and 2 are assigned to folds 5 and 1 respectively in this example. There are total five folds because \code{nfolds = 5}.

\begin{Verbatim}[fontsize = \normalsize]
> listdata = simuldata(n = 5000, l = 2, ep = -3)$data
> head(listdata, 3)
  y1 y2       x1       x2
1  1  0 2.159287 2.258739
2  0  1 2.654734 4.691390
3  0  1 5.338062 1.279576
> estim = popsize(data = listdata, funcname = c("rangerlogit"), nfolds = 5)
> head(estim$nuis)
  listpair rangerlogit.q12 rangerlogit.q1 rangerlogit.q2
1      1,2       0.1284399      0.7116891      0.4167508
2      1,2       0.2425309      0.7532815      0.4892493
3      1,2       0.2012161      0.8156479      0.3855682
4      1,2       0.2832612      0.8827187      0.4005426
5      1,2       0.4362669      0.8679311      0.5683358
6      1,2       0.2755165      0.8208690      0.4546476
> head(estim$idfold)
[1] 5 1 2 2 3 3

\end{Verbatim}

\begin{figure}[t]
    \centering
    \includegraphics[scale = 0.6]{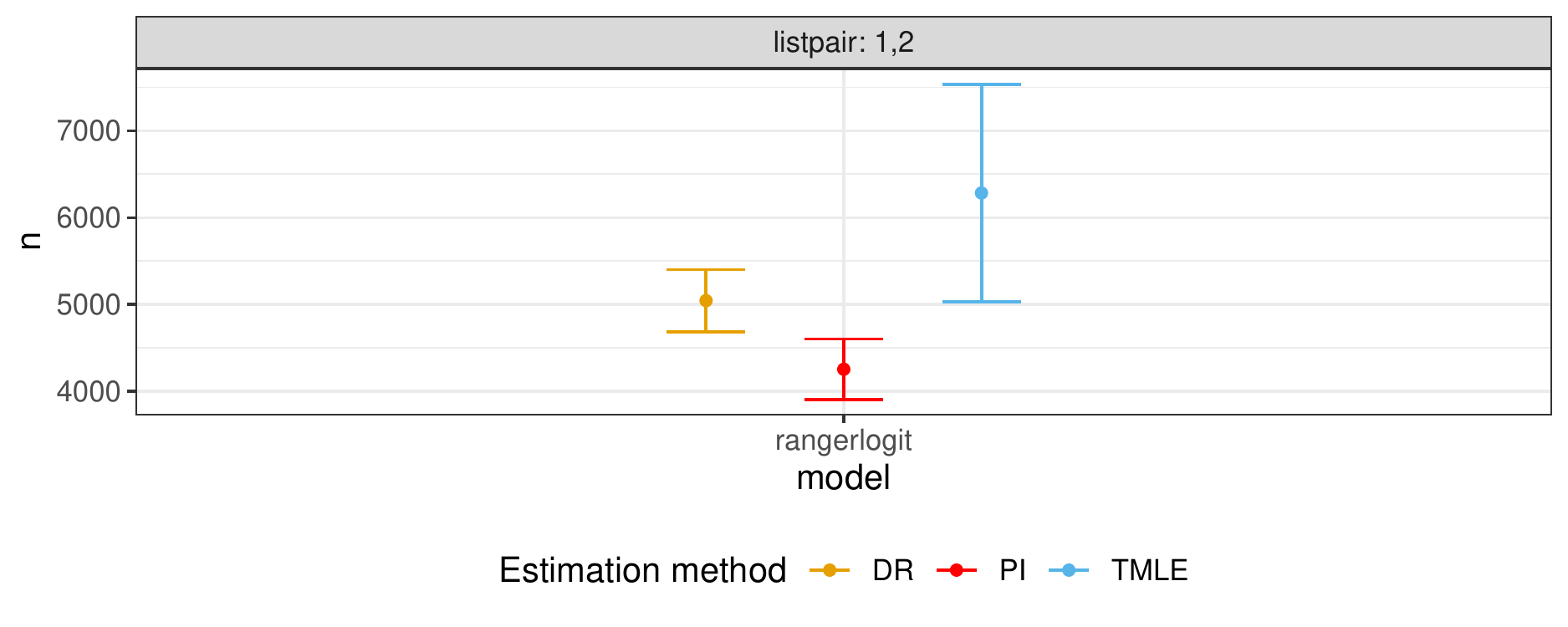}
    \caption{The plot above, generated by \code{plotci}, shows the estimated confidence interval for $n$ for the user provided nuisance estimates under two models, gam and logit. The variable list-pair (1,2) presents the conditionally independent lists. \code{popsize} returns results for only one list-pair which is the first list pair unless specified otherwise by the user.}
    \label{fig:cinnuis}
\end{figure}

Once we have the nuisance parameter estimates, we can pass it into \code{popsize}. As mentioned above, there are multiple ways of executing this. We illustrate the most straightforward approach below by passing \code{estim\$nuis} and \code{estim\$idfold} directly. The result is shown below and presented in Figure \ref{fig:cinnuis}. 

\begin{Verbatim}[fontsize = \normalsize]
> result = popsize(data = listdata, getnuis = estim$nuis, idfold = estim$idfold)

>result
  listpair       model method   psi  sigma    n  sigman cin.l cin.u
1      1,2 rangerlogit     DR 0.500  3.355 5045 182.890  4686  5403
2      1,2 rangerlogit     PI 0.593  3.355 4255 176.999  3908  4602
3      1,2 rangerlogit   TMLE 0.402 12.549 6284 637.844  5034  7535
> plotci(result)
\end{Verbatim}

\begin{remark}
In the current version of the package, one can pass nuisance parameter estimates only for one list-pair at a time. The lists can be specified using \code{j} and \code{k}. The default is \code{j = 1} and \code{k = 2}.
\end{remark}

\begin{remark}
All the datasets used in the examples in this section are simulated data.
\end{remark}

The package has an in-built function \code{simuldata} to generate a toy dataset with two or three lists. It can be used to test models by comparing against the true value. The \code{simuldata} function takes in the number of lists (\code{K}, default value 2), the number of continuous covariates (\code{l}), the logical option to include one categorical column (\code{categorical}, default value \code{FALSE}) and a numeric parameter to control the capture probabilities (\code{ep}, default value 0). It returns the empirical capture probability (\code{psi0}), the simulated dataset (\code{data}), the simulated dataset with transformed continuous covariates (\code{data\_xstar}) and the list wise capture probability functions (\code{pi1}, \code{pi2}, \code{pi3}) depending on \code{K}. For example, the function \code{pi1} returns the probability of being observed in list 1 for the covariate vector passed into it. The dataset with transformed covariates, \code{data\_xstar} can be used to study the robustness of a model.

\section{Performance}\label{sec:perform}

To motivate the use of the doubly-robust estimators of {drpop} for closed population, we present some summary statistics of its performance in a simulated set-up. The main focus of {drpop} is to flexibly estimate the total population size and at the same time to achieve optimal $1/n$ mean squared errors. The identifiability assumption used in {drpop} requires just two lists which are independent conditional on the covariates. We use simulated data that roughly satisfies this assumption to measure the performance of the proposed method. First, we present a comparison of the proposed doubly robust estimator in {drpop} against the baseline plug-in estimator under the flexible nonparametric set-up and also when any of the covariates are not correctly specified. Following that, we present some performance comparison of the closed population set-ups of the packages {Rcapture}, {CARE1}, {VGAM} and {drpop}.

\subsection{Performance in simulated set-up}

In this section, we evaluate the performance of the proposed method in the package over a 100 iterations and different simulation set-ups. Our simulation set-up ensures that the two lists are independent conditional on the covariates. The goal is to compare the performance against the baseline plug-in estimator. Moreover, we also compare the robustness of the estimators when the covariates are not correctly specified or are transformed. For the later case, flexible non-parametric models would prove useful. {drpop} has the choice of several such models. But for this section, we only use the default model \code{rangerlogit} which is an ensemble of logit and random forest models.

We simulate a data-frame using the \code{simuldata} function for two lists. The true population size is 5000 for each iteration. We generate data with true capture probabilities 0.36 and 0.75 separately. One can set the parameter \code{ep} equal to -2.5 and -1 respectively for the same. The code used to generate the data is
\begin{Verbatim}
> datalist = simuldata(n = 5000, l = 1, ep = -2.5)
\end{Verbatim}
where \code{n} is the true population size and \code{l} is the number of continuous covariates. The default number of lists in two. One can access the simulated data using \code{datalist\$data} and \code{listdata\$data\_xstar} where the later data-frame contains transformed (misspecified covariates).

\begin{figure}[t]
    \centering
    \includegraphics[width = \textwidth]{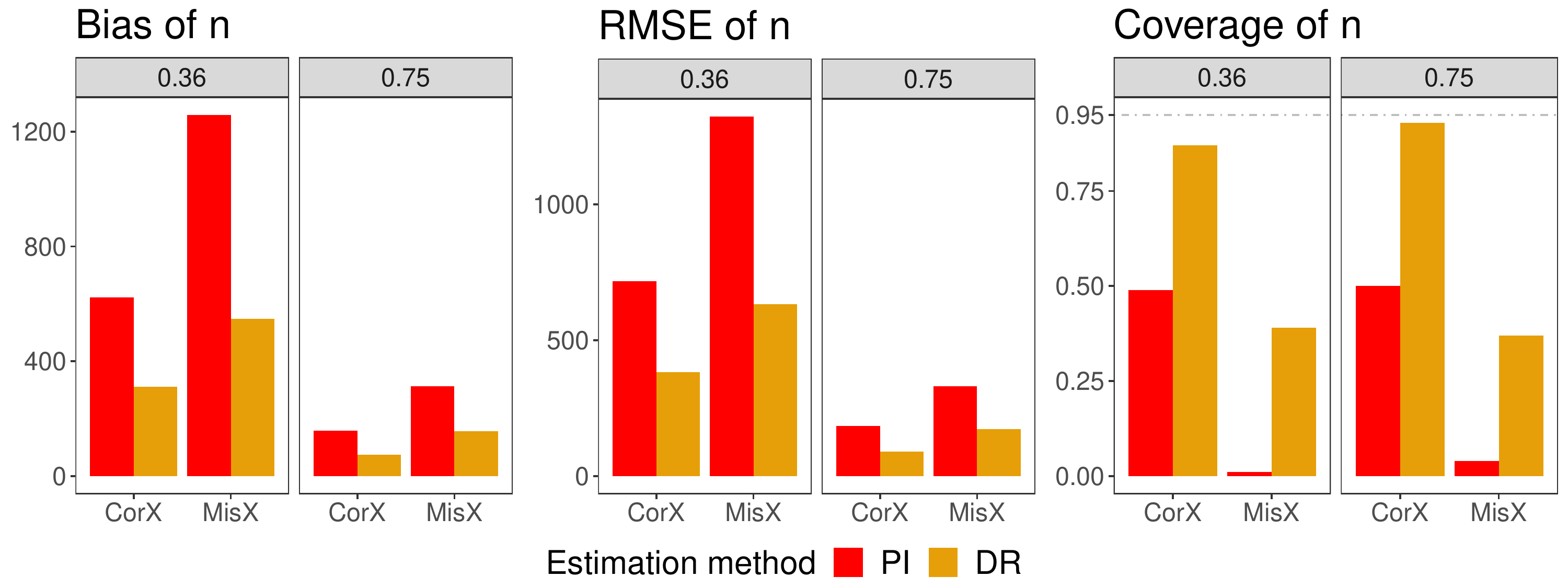}
    \caption{This figure shows the estimated average bias, root-mean-square-error (RMSE) and the empirical coverage in the estimation of the total population size \code{n}. The two facets show the true capture probability which is also the ($\times 100$)\% of the population observed. \code{CorX} and \code{MisX} refer to estimation using data with original covariates and transformed(mis-specified) covariates respectively. The doubly robust (\code{DR}) estimator has better performance in the set-up shown.}
    \label{fig:simul_barplot}
\end{figure}

We evaluated the bias, the RMSE (root-mean-square-error) and the empirical coverage by
\begin{align*}
    &\widehat{bias} = \frac{1}{100} \sum_{i=1}^{100} \left|\widehat{n}_i - 5000\right|^2, \quad \widehat{RMSE} = \sqrt{\frac{1}{100} \sum_{i=1}^{100} \left(\widehat{n}_i - 5000\right)^2},\\
    & \text{and} \ \widehat{coverage} = \frac{1}{100} \sum_{i=1}^{100} \one\left(\widehat{cin.l}_i \leq 5000 \leq \widehat{cin.u}_i\right),
\end{align*}
where $i$ is the iteration, $\widehat{n}_i$ is the estimated population size at iteration $i$, and $\widehat{cin.l}_i$ ($\widehat{cin.u}$) denote the estimated lower (upper) limit of the 95\% confidence interval. At each iteration, we generate a dataset independently of the other iterations. We evaluate these three quantities for both the capture probabilities, and also under the correct covariate data and misspecified covariate data. The results are shown in Figure \ref{fig:simul_barplot}. Overall, both methods perform better when we have a higher capture probability or the correct covariates. The proposed method (\code{DR}) has lower bias, lower RMSE and a higher empirical coverage compared to the plug-in \code{PI} estimator under both correct covariates and mis-specified covariates.

\subsection{Comparison to other packages}

We present some comparisons with existing {R} packages that can work for closed populations. We use the functions \code{closedp}, \code{estN} and \code{vglm} from the packages {Rcapture}, {CARE1} and {VGAM} respectively. We have considered two set-ups: two list case and three list case. {CARE1} requires more than two lists for its sample coverage approach and hence, we drop this package in the two list case. {Rcapture} uses log-linear models as discussed in \citet{JSSv019i05} and does not use covariate information. It is designed to use information from many lists to model the heterogeneity. {VGAM} uses log-likelihood approach and can incorporate continuous covariate information using generalized linear/additive models. We used simulated data to compare the performance of {drpop} against the models in the packages {Rcapture}, {CARE1}, and {VGAM} in a closed population set-up in the following two subsections. We note that these packages are based on assumptions different from ours. We present the comparison result for the sake of completeness.

\subsubsection{Two list case}

We begin with the simple case where we have only $K=2$ lists with some covariate information. The data is simulated using the simuldata function with parameters \code{K=2, l=1, ep=-1.5} i.e., the covariate is of dimension one. The total population size takes values in (3,000, 6,000, 9,000, 12,000, 15,000) and the true capture probability is approximately 0.63.

\begin{figure}[ht]
    \centering
    \includegraphics[width = \textwidth]{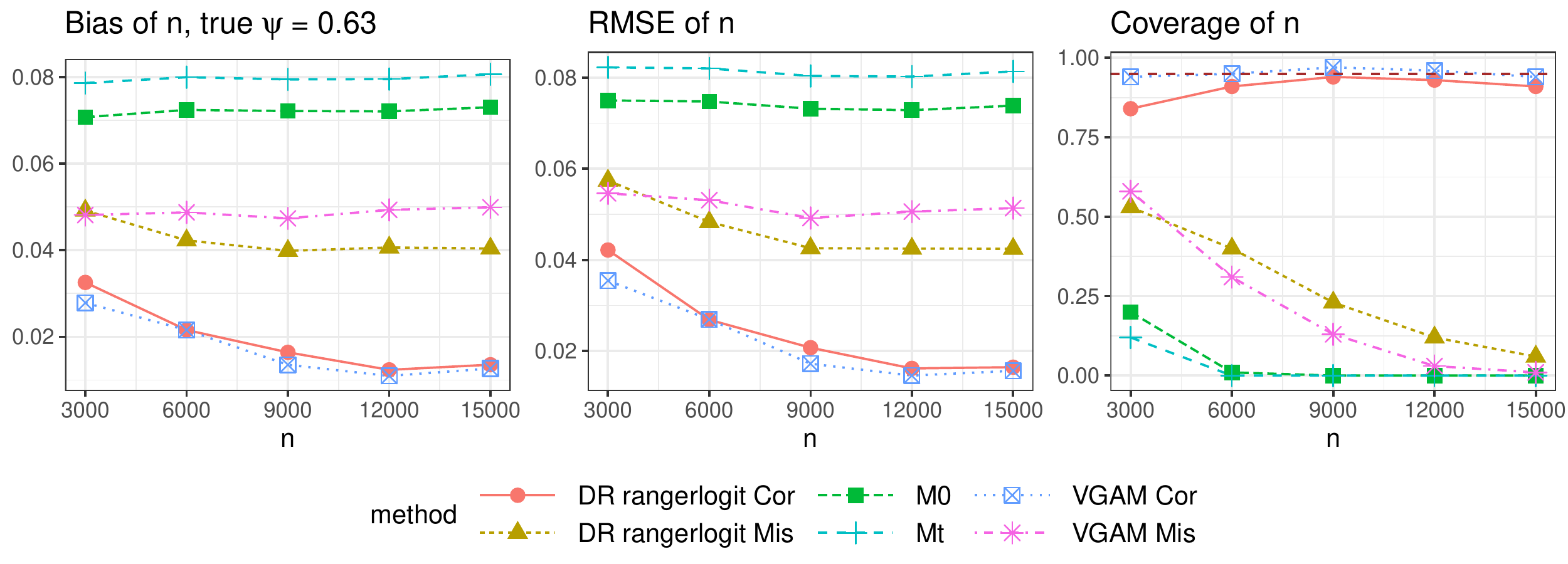}
    \caption{The absolute bias, root mean square error (RMSE) scaled by the true $n$, and empirical coverage of $n$ from the default model in {drpop} (\code{DR rangerlogit}), the \code{M0} and \code{Mt} models from {Rcapture} and the model in {VGAM}. The true population sizes are shown on the x-axis and the the true capture probability is 0.63, i.e., we are observing around 63\% of the population. For {drpop} and {VGAM} we present results with correctly specified covariates (Cor) and transformed/mis-specified covariates (Mis).}
    \label{fig:compare_Rcapture}
\end{figure}

{Rcapture} function \code{closedp} only fits three models (\code{M0} for no henerogeneity, \code{Mt} for list heterogeneity and \code{Mb} for heterogeneity based on first capture) when there are only two lists. For a full list of models, one can refer to \citet{JSSv019i05}. For \code{vglm} from package {VGAM}, we used posbernoulli.t to include list and individual heterogeneity. For {drpop}, we used the \code{rangerlogit} model to calculate the doubly robust estimator. Both {drpop} and {VGAM} use covariate information. Hence, we further compare their performance in terms of robustness of errors in covariate information i.e., transformed covariates. We applied them on data with the correctly specified/original covariates, and then on data with transformed/mis-specified covariates as in Figure \ref{fig:simul_barplot}. The results are presented in Figure \ref{fig:compare_Rcapture}.

We removed the estimates of the \code{Mb} model from the plot because, it had significantly large errors compared to the other methods (this is expected based on the simulation set-up). In the above two list set-up, the estimate using the {drpop} and {VGAM} packages have bias and RMSE decreasing with the total population size at a faster rate compared to models \code{M0} and \code{Mt}. The coverage of the estimated confidence intervals is also closer to the nominal level of 95\% when the covariates are correctly specified. {VGAM} has slightly better coverage when the covariates are correctly specified. This is a consequence of the simulation set-up where the actual list probabilities are additive functions of the covariates. However, for the mis-specified covariates, {drpop} has slightly better performance for larger sample sizes.

\begin{remark}
The performance result in Figure \ref{fig:compare_Rcapture} is not necessarily a general phenomenon. This can change based on the simulation set-up, for example. More exploration is needed to figure out if this is general.
\end{remark}

\subsubsection{Three list case}

In this section, we apply our method and functions from the three packages in a three list set-up ($K = 3$). Our goal in this section is to show that the performance of {drpop} with the default parameter values, at least matches the performance of {Rcapture}, {CARE1} and {VGAM}. We again note that these packages are developed based on assumptions different than those of package {drpop}. Hence, we do not expect unbiased estimates.

We use \code{simuldata} function to generate toy population with three lists and three dimensional continuous covariates. We set \code{ep} at -5 and -3 to get true capture probability \code{psi0} equal to 0.34 and 0.80 respectively. We set the true total population size at 15,000 and 5,000 for 0.34 and 0.80 respectively, since a low capture probability requires a larger number of observations for good estimation. The number of observations for the two set-ups are approximately 5,100 and 4,000 for each iteration. For the above set-up we generated a simulated dataset 100 times and estimated the population size for each.

\begin{figure}[t]
    \centering
    \includegraphics[width = 3.4in, height = 2 in]{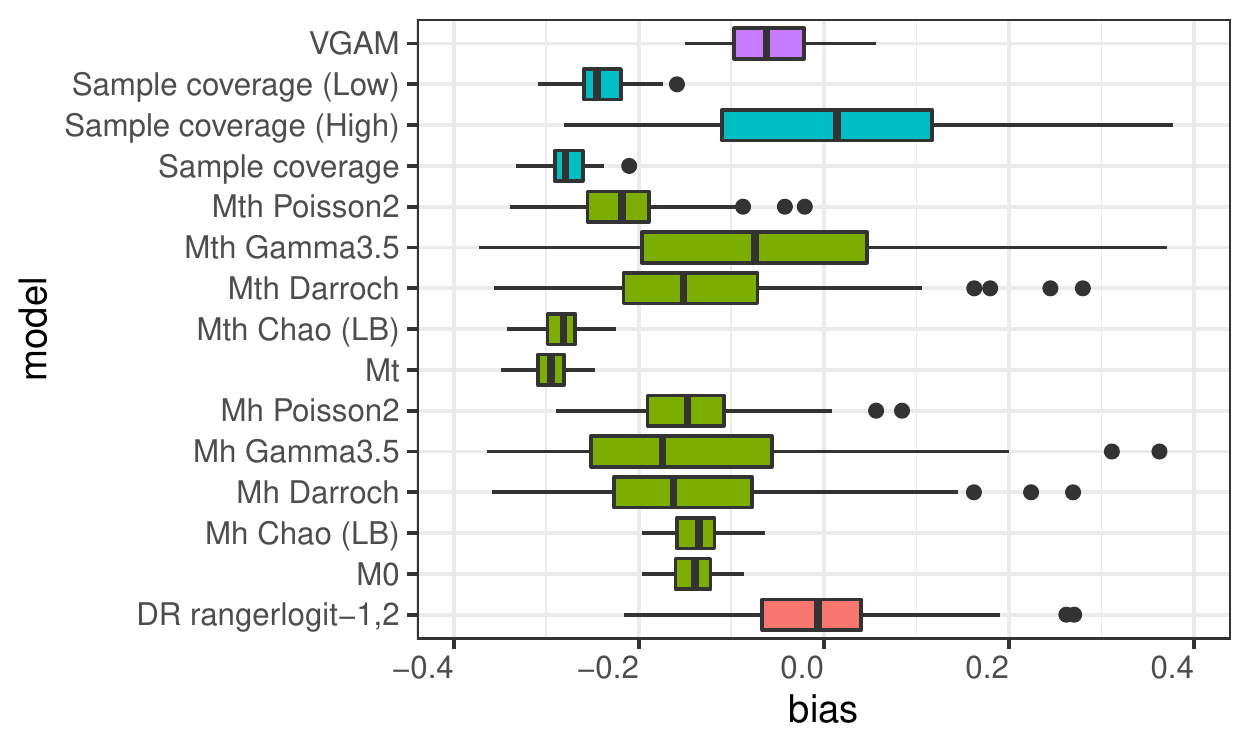}\includegraphics[width = 2.38in, height = 2 in, trim = {1.65in 0 0 0}, clip]{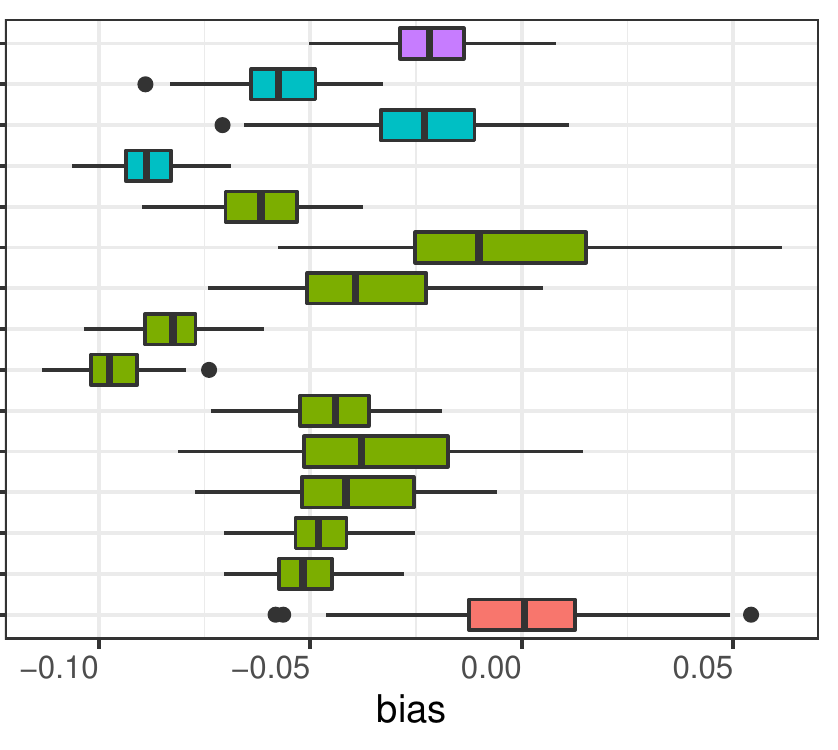}
    \caption{Scaled bias $(\widehat{n}-n)/n$ in the estimation of the total population size using four different packages: violet ({VGAM}), teal ({CARE1}), green ({Rcapture}) and red ({drpop}). {CARE1} and {Rcapture} return multiple estimates. On the left we observe 34\% of the whole data, and on the right we observe 80\% of the data.}
    \label{fig:compare_3list}
\end{figure}

To compare the performance, we present the boxplot of the scaled bias $(\widehat{n} - n)/n$ of each iteration for the different models in Figure \ref{fig:compare_3list}. The estimation models from the four packages are marked by colors. The doubly robust estimator is \code{DR rangerlogit} using only the first two lists for simplicity. For {Rcapture}, we excluded the \code{Mb} and \code{Mbh} models because they have large error which is expected under the current simulation set-up. All the estimators display better performance (lower bias and/or lower variance) for capture probability 0.8 compared to 0.34. For the specific set-up used with 80\% observed data, the proposed method and \code{Mth Gamma3.5} have bias closest to 0 followed by \code{VGAM} and \code{Sample coverage (High)}. Whereas, for the 34\% observed data set-up, the proposed method and \code{Sample coverage (High)} have bias closest to 0 followed by \code{VGAM} and method \code{Mth Gamma3.5}.

\begin{remark}
The performance result in Figure \ref{fig:compare_3list} is not necessarily a general phenomenon. More exploration is needed to figure out if that is the case.
\end{remark}

Summarizing the results and the advantages of {drpop}, it is capable of incorporating high dimensional and complex covariates as well as interaction among the covariates. The user can choose from several flexible modelling options that are provided in the package or also, use their own models to estimate the nuisance parameters. Under the identification assumption of conditional independence between two lists, the proposed estimator in {drpop} also handles mis-specified covariates better compared to the naive plug-in estimator. Further, attributed to the bias-correction step, the estimation under small capture probability (small observed sample) is also better compared to the plug-in estimator.

\section{Discussion}

In this paper, we have presented the {R} package {drpop} to implement a new doubly robust estimator of the total population size and an associated confidence interval from incomplete lists.  The package provides  users with many choices for flexibly modelling the heterogeneity which usually exists in real data. Further, the proposed method implemented in the package \citep{das2021doubly} exploits efficiency theory so that it achieves beneficial properties such as (i) $1/n$ mean squared errors even in flexible nonparametric models, (ii) double robustness, (iii) minimax optimality, and (iv) near finite-sample normality. 

One of the main advantages of {drpop} is that it can model the heterogeneity in the data as complex functions of discrete and/or continuous covariates. This is useful when the capture probabilities (nuisance parameters i.e. $q_1$, $q_2$, $q_{12}$) of the individuals do not depend linearly on the covariates. More discussion on this can be found in \citet{yee1991generalized, crawley1993glim, gimenez2006nonparametric, bolker2008ecological, schluter1988estimating}, and \citet{yee2015vgam}. \citet{yee2015vgam} also created an {R} package {VGAM} which addresses this issue via vector generalized models. The availability of flexible models in {drpop} makes it easy for users to obtain good estimates for such datasets as well. The users also have the option to fit their own models to estimate nuisance parameters and pass them into the package functions to obtain total population size estimate and confidence interval(s).

The estimation method implemented in {drpop} exploits modern advances in nonparametric efficiency theory. This ensures that even when one is using flexible nonparametric methods, the rate of convergence (i.e., mean squared error) is not compromised. Typically, plug-in estimators inherit  convergence rates from the estimators of the more complex nuisance parameters like $q$-probabilities. However, because of the form of the proposed estimator, we can still achieve  $1/n$ mean squared errors, even when the nuisance functions are estimated flexibly at slower rates. Further, the estimate is doubly robust against errors in the estimation of the nuisance parameters. In particular, even when either one of $q_1$ and $q_2$ is estimated with large errors, or one of $q_{12}$ and $\gamma$ is estimated with large errors, the $\widehat\psi$ and $n$ still have bounded errors as long as $q_{12}$ and $\widehat{q}_{12}$ are bounded away from zero. More details and explanation can be found in \citet{das2021doubly}. Further, as a consequence of efficiency theory, this estimator is near minimax optimal in finite samples and has a nearly normal distribution, permitting simple but valid  confidence interval construction.

We have presented some simulation results in section \ref{sec:perform} to show the advantages of the proposed estimator against the baseline method. We have also provided some simulation results to compare the performance of the proposed method in {drpop} against some of the existing widely used {R} packages for the closed population set-up. Our goal is to show that when the capture probability depends on covariates and when our mild identifiability assumption holds, the performance of {drpop} is reliable and comparable to some of the existing methods for the given set-up. 

Alongside the proposed doubly robust estimator, this package also provides the user with the choice of evaluating the baseline plug-in estimator and an alternate targeted maximum likelihood estimator (TMLE). Some of the other functions this package can perform are (i) simulate toy data for model training and study design, (ii) estimate total population size and other parameters and other information for sub-populations based on a categorical covariate, and (iii) plot the results with an in-built function for easy and fast interpretation. A full list is presented in section \ref{sec:implement}.

\newpage
\appendix

\newpage
\bibliographystyle{plainnat}

\end{document}